\documentclass[a4paper,12pt]{article}
\usepackage{amsmath}
\usepackage{amssymb}
\usepackage{amsfonts}
\usepackage{graphics}
\usepackage{epsfig}
\usepackage{epic}
\usepackage[latin1]{inputenc}

\begin{document}

\renewcommand{\theequation}{\arabic{section}.\arabic{equation}}
\thispagestyle{empty}
\vspace*{-1,5cm}
\hfill{}
 \\[8mm]

\begin{center}
{\large THE GOLDSTONE FIELDS OF INTERACTING HIGHER \\ SPIN FIELD THEORY IN ADS(4)}\\
\vspace{2cm}
{\large Werner Rühl}\\
Department of Physics, Kaiserslautern University of Technology\\
P.O.Box 3049 \\
67653 Kaiserslautern, Germany \\
\vspace{5cm}
\begin{abstract}
A higher spin field theory on AdS(4) possesses a conformal theory on the boundary R(3)
which can be identified with the critical O(N) sigma model of O(N) invariant fields only.
The notions of quasiprimary and secondary fields can be carried over to the AdS theory.
If de Donder's gauge is applied, the traceless part of the higher spin field on AdS(4) is 
quasiprimary and the Goldstone fields are quasiprimary fields to leading order, too. Those 
fields corresponding to the Goldstone fields in  the critical O(N) sigma model are odd rank symmetric
tensor currents which vanish in the free field limit.
\end{abstract}
\vspace{5cm}
{\it July 2006}
\end{center}
\newpage

\section{Introduction}
If we move from the free field limit $N = \infty $ of the critical $O(N)$ sigma model to the interacting
conformal field theory $CFT(3)$ by a renormalization group transformation \cite{Kl,Le}, gauge symmetry of the higher spin field theory $HS(4)$ on the $AdS(4)$ side is broken and a Goldstone field arises. Such Goldstone modes are proper dynamical degrees of freedom  
that must be represented also in the boundary conformal field theory. Its quantum numbers being known,
we must try to find such field in the list of fields for $CFT(3)$.

To derive such list one looks first for all quasiprimary fields in the free field limit $N=\infty$.  
A simple algorithm tells us \cite{LR} how to choose the quasiprimary fields from all composites of the
derivatives of the free massless fields and select the $O(N)$ invariant ones from them. Symmetric traceless tensor fields are then always of even rank (odd rank tensors belong to nontrivial $O(N)$ 
representation and are eliminated). Then one perturbs these composites by switching on the interaction. 
They obtain this way anomalous dimensions. Several quasiprimary fields in the free theory have degenerate quantum numbers (tensor type and dimension). Certain linear combinations of them form eigenvectors with respect to the anomalous dimensions as eigenvalues. These are the interacting quasiprimary fields.

Among these quasiprimary fields in the free field theory are exceptional elementary representations
which belong to conserved currents, their divergence is zero. If we switch on the interaction, these 
currents obtain not only a (positive) anomalous dimension, their divergence is nonzero and gives a new 
quasiprimary field. Their two-point function is proportional to their anomalous dimension. In this fashion
we can obtain symmetric traceless tensor fields with odd tensor rank. The Goldstone degree of freedom 
on the boundary of $AdS_{4}$ is represented by these field operators (in de Donder's gauge). 
Thus the renormalization group which looks continuous, produces a discontinuous jump in the field theory 
Hilbert space when we leave the free field point.

In this article we summarize the relevant properties of the $O(N)$ invariant critical sigma model (section $2$). In particular this section is devoted to the currents in this field theory. Section $3$ deals with the Goldstone field in higher spin field theory $HS(4)$. In this section we introduce the concept of quasiprimary and secondary fields in $HS(4)$. Using this we can identify the Goldstone field both in $CFT(3)$ and in $HS(4)$. In the latter case its two-point function is given relying on $AdS/CFT$ correspondence. We close in section $4$ with some remarks on representation theory. 

At present the status of $AdS/CFT$ correspondence is that all field correspondences predicted have
been established, but an access to the anomalous dimensions (equivalent to the masses) of the quasi-primary fields of $HS(4)$ is still lacking.

\section{The O(N) invariant critical sigma model}
We start summarizing the stable conformal sigma model. Its basic field is a scalar O(N)-vector field
$\vec{\varphi}(x)$ and an O(N)-scalar field $\alpha(x)$ with the Lagrangian
\begin{equation}
\frac{1}{2}\int d^{3}x \partial_{\mu}\vec{\varphi}(x)\partial^{\mu} \vec{\varphi}(x) \\
+ z^{\frac{1}{2}}_{c}\int d^{3}x \alpha(x)\vec{\varphi}(x)\vec{\varphi}(x)
\label{2.1}
\end{equation}
The auxiliary field $\alpha(x)$ serves as a Lagrangian multiplier for the constraint
\begin{equation}
\vec{\varphi}(x)\vec{\varphi}(x) = \textnormal{constant}
\label{2.2}
\end{equation}

All conformal fields in a CFT can be obtained by an operator product expansion of appropriate fields, and any field is either conformal (quasiprimary) or a derivative of a conformal (secondary) field. The latter fields can be ordered in classes
according to the number $n$ of derivations. An element of the conformal group maps a field of CFT into another one
by a coordinate change
\begin{equation}
x \longrightarrow \Lambda x
\label{2.3}
\end{equation}
and a multiplier depending on $x$ in general. Therefore a derivative field transforms such that derivatives of the
multiplier appear besides the multiplier itself. On the other hand a local field on AdS(4) is covariant with respect to
the isometry group of AdS space which is the same conformal group as in CFT but the multipliers are independent of the coordinate $w$ in the conventional representation. The coordinate transforms as
\begin{equation}
w \longrightarrow \tilde\Lambda w
\label{2.4}
\end{equation}
However, the difference between quasiprimary and secondary fields on AdS is visible in their two-point functions.

We are interested in the O(N) invariant subCFT generated by $\alpha(x)$ and the bilocal field \cite{R1}
\begin{equation}
b(x_1, x_2) = N^{-\frac{1}{2}}\vec{\varphi}(x_1)\vec{\varphi}(x_2)
\label{2.5}
\end{equation}
We normalize fields by ($x_{12} = x_1 - x_2$)
\begin{eqnarray}
<\alpha(x_1)\alpha(x_2)>_{\mathtt{CFT}} &=&  (x_{12}^{2})^{-\delta(\alpha)}\\
\label{2.6}
<\varphi_i(x_1)\varphi_j(x_2)>_{\mathtt{CFT}}  &=&  \delta_{ij}(x_{12}^{2})^{-\delta(\varphi)}
\label{2.7}
\end{eqnarray}
where the field dimensions $\delta$ are
\begin{eqnarray}
\delta(\alpha) &=& 2 + \eta(\alpha)\\
\label{2.8}
\delta (\varphi) &=& \frac{d}{2} - 1 + \eta(\varphi)
\label{2.9}
\end{eqnarray}
and the anomalous parts $\eta$ are expandable as
\begin{equation}
\eta = \sum_{r=1}^{\infty}\frac{\eta_{r}}{N^{r}}
\label{2.10}
\end{equation}
Only for a few leading powers the coefficients $\eta_{r}$ are known. We shall give them only for the
fields of interest in the present context. Moreover the critical coupling constant can be expanded in the
same way
\begin{equation}
z_{c} = \sum_{r=1}^{\infty} \frac{z_{r}}{N^{r}}
\label{2.11}
\end{equation}
with
\begin{equation}
z_{1} = \frac{1}{\pi^{4}}
\label{2.12}
\end{equation}

The even rank currents that tend to conserved currents in the free field limit can be derived by operator product expansion (which for four-point functions is almost the same as a conformal partial wave expansion) of the four-point function
\begin{equation}
<b(x_{1}, x_{3}) b(x_{2},x_{4})>_{\mathtt{CFT}}
\label{2.13}
\end{equation}
in the limit
\begin{equation}
x_{13} \rightarrow 0, x_{24} \rightarrow 0
\label{2.14}
\end{equation}
The $\frac{1}{N}$ expansion of (2.13) can be obtained from a graphical expansion so that to each order $\frac{1}{N^{r}}$
there contributes a finite number of graphs. This has been done in the literature a long time ago \cite{RL}.
The order $r = 0$ can be evaluated exactly and is most conveniently expressed in terms of variables
\begin{eqnarray}
u = \frac {x_{13}^{2}x_{24}^{2}}{x_{12}^{2}x_{34}^{2}},\qquad
\label{2.15}
v = \frac {x_{14}^{2}x_{23}^{2}}{x_{12}^{2}x_{34}^{2}}
\label{2.16}
\end{eqnarray}
as
\begin{equation}
(x_{12}^{2}x_{34}^{2})^{-\delta(\varphi)}\{ 1 + v^{-\delta(\varphi)} + C_{1}F_{\alpha}(u, v)
+ C_{2}F_{\vec{\varphi}^{2}}(u, v)\}
\label{2.17}
\end{equation}

The first two terms arise from the disconnected graphs and the last two terms from an exchange graph of the $\alpha$ field in
the channel $(1,3)\rightarrow (2,4)$. Both functions F are series in $u$ and $1-v$. Whereas $F_{\alpha}$ represents a single
conformal partial wave for the representation of the field $\alpha$, the remaining three terms give by harmonic analysis
conformal partial waves for the currents $j^{l}(x)$ for all $l\in 2\mathbf{N}$ and no other fields \cite{R1}. That $l= 0$ is excluded follows from the fact that
\begin{equation}
1 + v^{-\delta(\varphi)} + C_{2}F_{\vec{\varphi}^{2}} = 0 \qquad\textnormal{for}\qquad u = 1-v = 0
\label{2.18}
\end{equation}
The currents are traceless symmetric tensors of even rank (spin) $l$ and at this order (i.e. $r = 0$) of $1/N$ they are conserved with dimension 
\begin{equation}
\delta(j^{l}) = d + l - 2 \qquad( = l + 1 \qquad\textnormal{at}\qquad d = 3)
\label{2.19}
\end{equation}

At the next order $1/N$ the four-point function (2.13) involves box graphs which can still be analyzed. One group
of terms ($logu$ terms) correct the dimensions of the fields found at leading order \cite {Vas}, \cite {R1}
\begin{eqnarray}
\delta(\alpha) &=& 2 + \eta(\alpha), \qquad\eta_{1}(\alpha) = \frac{16}{3\pi^{2}}\\
\label{2.20}
\delta(j^{l}) &=& l + 1 +\eta(j^{l}), \qquad\eta_{1}(j^{l}) = \frac{16(l-2)}{3\pi^{2}(2l-1)}
\label{2.21}
\end{eqnarray}
The currents $j^{l}$ are no longer conserved except for $l = 2$.

At the same order another group of terms can be analyzed in terms of conformal partial waves for the representations
of the "twist currents" $j^{l,t}(x), l\in 2\mathbf{N}, t\in\mathbf{N}$ with dimensions at leading order
\begin{equation}
\delta(j^{l,t}) =  d + l - 2 + 2t\qquad (= l + 1 + 2t \qquad\textnormal{at}\qquad d = 3)
\label{2.22}
\end{equation}
which are also traceless symmetric tensors of even rank (spin) $l$. They are never conserved.

In order to determine their anomalous dimensions $\eta(j^{l,t})$ one has to go to the order $\frac{1}{N^{2}}$ and extract the $log u$ terms. This is rather involved and has not been done yet. It may be simpler to start from a $(2t+4)$-point function,
since these operators can also be obtained as normal products of
\begin{equation}
j^{l}(x_{1})\alpha(x_{2})\alpha(x_{3})\cdots \alpha(x_{t+1})
\label{2.23}
\end{equation}

For the later applications we need some technical tools. Contracting the tensor $j^{l}(x)$ with $l$ $d$-vectors $a$ ($d=3$ of
course), the result is denoted $j^{l}(x;a)$. From representation theory we know that its two-point function is
\begin{equation}
<j^{l}(x_{1};a) j^{l}(x_{2};b)>_{\mathtt{CFT}} =  \mathcal{N}_{l} (x_{12}^{2})^{-\delta(j^{l})}\\
\{[(ab)- 2\frac{(ax_{12})(bx_{12})}{x_{12}^{2}}]^{l} - \textnormal{trace terms}\}
\label{2.24}
\end{equation}
The subtraction of traces in (2.23) is a power expansion in $a^{2}b^{2}$ and tracelessness is the same as harmonicity with respect to the Laplacians $\triangle_{a}$ or $\triangle_{b}$. Therefore we can express the formula (2.23) by a
Gegenbauer polynomial
\begin{eqnarray}
&b' = b - 2\frac{(bx_{12})}{x_{12}^{2}}x_{12},\qquad  (b')^{2} = b^{2}\\
\label{2.25}
&<j^{l}(x_{1};a) j^{l}(x_{2};b)>_{\mathtt{CFT}} = \mathcal{N}_{l}(x_{12}^{2})^{-\delta(j^{l})}
(a^{2}b^{2})^\frac{l}{2} \frac{l!}{2^{l}(\frac{d}{2}- 1)_{l}} C_{l}^{\frac{d}{2}-1}(\frac{ab'}{(a^{2}b^{2})^{1/2}})\\
\label{2.26}
  &= \mathcal{N}_{l}(x_{12}^{2})^{-\delta(j^{l})} (ab')^{l}  \cdot_{2}F_{1}(-\frac{l}{2},\frac{1-l}{2};2-l-\frac{d}{2};\frac{a^{2}b^{2}}{(ab')^{2}})
\label{2.27}
\end{eqnarray}
(see \cite{RG}).

In the $O(N)$ invariant subCFT of the conformal sigma model, which we can call the "minimal" sigma model,
and after a normalization of all basic fields, $N$ is a positive free parameter that we can assume to vary over the reals.
Because we handle this model with a 1/N perturbative expansion, maybe only large $N$ are relevant. By $AdS/CFT$
this model is mapped on a local $AdS$ field theory with coupling constant $z_{c}(N)$. Relevant are only small
coupling constants, and these appear in an infinite number of coupling terms. Since the sigma model is renormalizable,
it seems that this field theory $HS(4)$ is also well defined. However, the correspondence may be
only classical and not apply directly to n-point functions including non-tree graphs.

\setcounter{equation}{0}
\section{The Goldstone field in HS(4)}
We define the higher spin field $h^{l}(w)$ on AdS(4) as a symmetric double-traceless tensor field with respect to the tangential planes of AdS(4) and contract it therefore with $l$ $(d+1)$-vectors $a$ resulting in
\begin{equation}
h^{l}(w;a): \Box_{a}\Box_{a} h^{l}(w;a) = 0
\label{4.1}
\end{equation}
We decompose $h^{l}$ into two traceless fields \cite{MR1}
\begin{equation}
h^{l}(w;a) = \psi^{l}(w;a) + \frac{a^{2}}{2\alpha_{0}}\theta^{l-2}(w;a)
\label{4.2}
\end{equation}
where
\begin{equation}
\alpha_{0} = d + 2l - 3 \quad (= 2l\quad  \textnormal{at}\quad d = 3)
\label{4.3}
\end {equation}
so that
\begin{eqnarray}
\Box_{a}\psi^{l}(w;a) = 0\\
\label{4.4}
\Box_{a}h^{l}(w;a) = \theta^{l-2}(w;a)
\label{4.5}
\end{eqnarray}
Thus under the isometry group $\psi^{l}$ transforms as $[\Delta, l]$ with $\Delta = \delta(j^{l})$ (2.18),
(2.20) and $\theta^{l-2}$ as $[\Delta, l-2]$. Consequently we have assuming AdS/CFT correspondence to be valid
\begin{equation}
\lim_{w_{10}\to 0,w_{20}\to 0} (w_{10}w_{20})^{-\Delta +l}<\psi^{l}(w_{1}; a)\psi^{l}(w_{2};b)>|_{a_{0} = b_{0}=0}  =  <j^{l}(x_{1}; a')j^{l}(x_{2};b')>_{\mathtt{CFT}}
\label{4.6}
\end{equation}

In free field theory $h^{l}(w;a)$ satisfies Fronsdal's equation  \cite{F}
\begin{eqnarray}
\lefteqn{\mathcal{F}(h^{l}(w;a)) \equiv \Box h^{l}(w;a) - (a\nabla)\nabla^{\mu}\frac{\partial}{\partial a^{\mu}}h^{l}(w;a) + \frac{1}{2} (a\nabla)^{2}\Box_{a}h^{l}(w;a) + {} }\nonumber\\
& & {} + (l^{2} - 3l +(l-2)(d-2))h^{l}(w;a) + a^{2}\Box_{a}h^{l}(w;a)  = 0
\label{4.7}
\end{eqnarray}
By choosing de Donder's gauge we can separate $h^{l}$ and $\theta^{l-2}$ in (3.7). In fact
\begin{equation}
\nabla^{\mu}\frac{\partial}{\partial a^{\mu}} h^{l} - \frac{1}{2}(a\nabla)\Box_{a}h^{l} = 0
\label{4.8}
\end{equation}
implies
\begin{equation}
\mathcal{F}_{dD}(h^{l}(w;a)) \equiv  \Box h^{l}(w;a) + (l^{2} - 3l + (l-2)(d-2))h^{l}(w;a) + a^{2}\Box_{a}h^{l}(w;a) = 0
\label{4.9}
\end{equation}
The Goldstone field is defined as a traceless tensor field of rank $l-1$ by \cite{MR2}
\begin{equation}
G^{l-1}(w;a) = (a\nabla)\theta^{l-2}(w;a) - \textnormal{traces}
\label{4.10}
\end{equation}
and belongs to the representation $[d+l-1+\eta(j^{l}), l-1]$ of the isometry group.

If we switch on the interaction and move to the critical coupling, the anomalous dimension $\eta(j^{l})$ produces a mass of $h^{l}$ by the formula
\begin{equation}
m_{l}^{2} = \Delta(\Delta -d) -(l-2)(d+l-2) = \eta(j^{l})(d+2l-4) + \eta(j^{l})^{2}
\label{4.11}
\end{equation}
Thus to leading order in 1/N the mass of $h^{l}$ is for $d=3$ \cite{R1} 
\begin{equation}
m_{l}^{2} = \frac{16(l-2)}{3\pi^{2}N}
\label{4.12}
\end{equation}
which vanishes for the graviton $l=2$.

We assume that all coupling terms are formally gauge invariant. Stueckelberg's Lagrangian formalism permits to render
the mass terms also gauge invariant by choosing the Goldstone fields as dynamically independent. We gauge transform
the Goldstone fields by \cite{MR2}
\begin{equation}
\delta G^{l-1}(w;a) = - \epsilon^{l-1}(w;a)
\label{4.13}
\end{equation}
corresponding to 
\begin{equation}
\delta h^{l}(w;a) = a\nabla \epsilon^{l-1}(w;a)
\label{4.14}
\end{equation}
and introduce the mass term actions
\begin{equation}
S_{m_{l}}[h^{l}, G^{l-1}] = \frac{1}{2} m_{l}^{2}\int dw \sqrt{-g}(h^{l} + \nabla G^{l-1})^{2}
\label{4.15}
\end{equation}
($\nabla G^{l-1}$ is symmetrized). We impose de Donder's gauge by (3.8) also for the massive fields.

We have learnt in \cite{MR1} that the Goldstone field projected on CFT(3) vanishes in the free field limit. But in the interacting case it survives. First we notice that the representation $[d+l-1+\eta(j^{l}), l-1]$ of $(a\nabla) \theta^{l-2}(x;a)$ is equal to the representation of the divergence of the current $j^{l}(x;a)$
and this equality includes the anomalous parts of the dimensions. 
In de Donder's gauge and in the free field limit the two-point function of the gauge field $h^{l}$ is
(from now on we use the bitensorial basis $I_{1}, I_{2}$ e.g. defined in \cite{MR2}, these are linear in $a$ and in $b$; $\zeta -1$ is the square of the chordal distance between the points $w_{1}, w_{2}$.)
\begin{equation}
\Psi^{l}[F] = (I_{1} - \frac{1}{\zeta} I_{2})^{l} \zeta^{-\Delta} F_{0}(\zeta) + \sum_{k=1}^{l} I_{1}^{l-k}I_{2}^{k}\zeta ^{-\Delta -k -1}f_{k}(\zeta)
            + \textnormal{trace terms}
\label{4.16}
\end{equation}
Without limiting the theory we can assume that $F_{0}(\zeta)$ and all $f_{k}(\zeta)$ are analytic functions of
$1/\zeta$ at $\zeta= \infty$, and that
\begin{equation}
F_{0}(\infty) \not= 0, \quad f_{k}(\infty)\quad \textnormal{finite}
\label{4.17}
\end{equation}
This was proven in \cite{MR1}.
Since the boundary limit of (4.16)
\begin{equation}
\lim_{w_{10}, w_{20}\rightarrow 0}(w_{10}w_{20})^{-\Delta + l}\quad \Psi[F](w_{1},w_{2})|_{a_{0} =b_{0}=0}
\label{4.18}
\end{equation}
is the two-point function of the quasiprimary tensor field $j^{l}$ of CFT(3), we call any AdS field with
a two-point function satisfying (3.16), (3.17) a quasiprimary field, explicitly a field of the generation zero $[\Delta, l]_{(0)}$. Using Gegenbauer polynomials one can partially inlude the trace terms in (3.16)
by
\begin{equation}
\Psi^{l}[F] = (I_{1}-\zeta^{-1} I_{2})^{l}\cdot_{2}F_{1}(-\frac{l}{2}, \frac {1-l}{2};
-\frac{\alpha_{0}}{2}; \frac{\frac{\alpha_{0}+1}{\alpha_{0}} I_{4} - \zeta^{-2}I_{3}}{(I_{1}-\zeta^{-1}I_{2})^{2}}) \zeta^{-\Delta}F_{0}(\zeta)(1+O(\zeta ^{-1}))
\label{4.19}
\end{equation}
with $\alpha_{0}$ as in (3.3).

In the algebra of bitensors (see \cite{MR1}) we can apply a bigradient operation
\begin{equation}
(a\nabla_{1})(b\nabla_{2})\Psi^{l}[F] = \sum_{k=0}^{l+1} I_{1}^{l+1-k} I_{2}^{k}(GradGradF)_{k}
                               + \textnormal{trace terms} 
\label{4.20}
\end{equation}
where
\begin{equation}
(GradGradF)_{k} =F_{k}'' + 2(k+1)F_{k+1}' + (k+1)_{2}F_{k+2}
\label{4.21}
\end{equation}
If we insert
\begin{equation}
F_{k} = (-1)^{k}\binom{l}{k}\zeta^{-\Delta -k} 
\label{4.22}
\end{equation}
into
\begin{equation}
\sum_{k=0}^{l}F_{k} I_{1}^{l-k}I_{2}^{k}
\label{4.23}
\end{equation}
then we obtain the first term in (3.16) with $F_{0}(\zeta) = 1$. If we insert (3.22) into (3.20), (3.21) and
after summing over $k$ we obtain
\begin{equation}
- (\Delta + l)(I_{1} - \frac{1}{\zeta} I_{2})^{l}(I_{1} -\frac{\Delta + l + 1}{\zeta}I_{2})\zeta^{-\Delta - 1}
\label{4.24}
\end{equation}
This bitensor characterizes a derivative field of generation $one$ which we label by the symbol $[\Delta + 1, l + 1]_{(1)}$.
Thus secondary fields on AdS are classified by a tensor representation and a generation number. In general a bitensor of the structure
\begin{equation}
\Psi^{l+1}[F] = (I_{1} -\frac{1}{\zeta}I_{2})^{l}(I_{1} - \frac{\Delta + l + 1}{\zeta}I_{2})\zeta^{-\Delta - 1}F_{0}\\
+ \sum_{k=1}^{l+1}I_{1}^{l+1-k}I_{2}^{k}\zeta^{-\Delta-k-2}f_{k}(\zeta) + \textnormal{trace terms}
\label{4.25}
\end{equation}
with $F_{0}$ and $f_{k}$ having the same properties as before, defines the generation  $[\Delta + 1, l + 1]_{(1)}$.

We close this discussion with two remarks. The boundary limit of the bitensor (3.25) has the closely related form
\begin{equation}
(x_{12}^{2})^{-\Delta-1}[(ab) - 2\frac{(ax_{12})(bx_{12})}{x_{12}^{2}}]^{l}[ab - 2(\Delta + l + 1) \frac{(ax_{12})(bx_{12})}{x_{12}^{2}}] + \textnormal{trace terms}
\label{4.26}
\end{equation}
and is identical with the flat bigradient operation applied to the quasiprimary two-point function in CFT(3).
The bigradient operation works recursively and maps a generation $[\Delta, l]_{(n)}$ into the generation $[\Delta + 1, l + 1]_{(n+1)}$ in AdS field theory by
\begin{eqnarray}
\sum_{k=0}^{l} a_{k}I_{1}^{l-k} I_{2}^{k} \zeta ^{-\Delta - k} \Rightarrow\nonumber\\ \sum_{k=0}^{l+1}A_{k}I_{1}^{l+1-k}I_{2}^{k}\zeta^{-\Delta-k-1}
\label{4.27}
\end{eqnarray}
with
\begin{equation}
A_{k} =(\Delta+k-1)_{2} a_{k-1} - (2k+1)(\Delta+k) a_{k} + (k+1)^{2} a_{k+1}
\label{4.28}
\end{equation}

The analogous bidivergence operation is 
\begin{equation}
(\nabla_{1}\frac{\partial}{\partial a})(\nabla_{2}\frac{\partial}{\partial b}) \Psi^{l}[F]
=\sum_{k=0}^{l-1} I_{1}^{l-1-k}I_{2}^{k}(DivDivF)_{k} +\textnormal{trace terms}
\label{4.29}
\end{equation}
But since $(DivDivF)_{k}$ is not available from the literature we shall not present it here. 
Instead we will use only the corresponding flat space relation for the analysis of the Goldstone field
for which this is the most important tool.

De Donder's gauge condition is assumed to be unchanged in the interacting case and in terms of two-point functions leads by (3.8), (3.5) to
\begin{equation}
(\nabla_{1}\frac{\partial}{\partial a})(\nabla_{2}\frac{\partial}{\partial b})
< h^{l}(w_{1};a)h^{l}(w_{2};b)> =
1/4 (a\nabla_{1})(b\nabla_{2}) <\theta^{l-2}(w_{1};a)\theta^{l-2}(w_{2};b)>
\label{4.30}
\end{equation}
where the equality holds with anomalous dimensions inclusive. Subtracting traces on both sides using
(3.2) -(3.5) and (3.10) we get 
\begin{equation}
(\nabla_{1}\frac{\partial}{\partial a})(\nabla_{2}\frac{\partial}{\partial b})<\psi^{l}(w_{1};a)\psi^{l}(w_{2};b)>  =
 \frac{1}{4} <G^{l-1}(w_{1};a)G^{l-1}(w_{2};b)>
\label{4.31}
\end{equation}
Thus the two-point function of the Goldstone field is a bidivergence of the two-point function of a quasiprimary field.
It is easier now to continue the arguments on the CFT(3) side. The corresponding field of $G^{l-1}$ 
in CFT is denoted $g^{l-1}$.

We start from (3.31), (2.26)
\begin{eqnarray}
&& \frac{1}{4} <g^{l-1}(x_{1};a)g^{l-1}(x_{2};b)>_{\mathtt{CFT}} =  \nonumber\\
&&(\frac{\partial}{\partial a}\frac{\partial}{\partial x_{1}})
(\frac{\partial}{\partial b}\frac{\partial}{\partial x_{2}})
\mathcal{N}_{l} (x_{12}^{2})^{-\delta(j^{l})}(ab')^{l} \cdot _{2}F_{1}(-\frac{l}{2}, \frac{1-l}{2};
2-l-\frac{d}{2}); \frac{a^{2}b^{2}}{(ab')^{2}})
\label{4.32}
\end{eqnarray}
The r.h.s. gives after performing the four differentiations and including the first trace 
term from the Gegenbauer polynomial
\begin{equation}
2l \eta^{l}\mathcal{N}_{l}\{[ab - 2\frac{(ax_{12})(bx_{12})}{x_{12}^{2}}]^{l-2}(Aab -2B\frac{(ax_{12})(bx_{12})}
{x_{12}^{2}}) + \textnormal{trace terms}\} (x_{12}^{2})^{-\delta(j^{l}) -1}
\label{4.33}
\end{equation}
and with $\eta^{l} = \eta(j^{l})$
\begin{eqnarray}
A &=& d+l-1+2\eta^{l} - 2\frac{l-1}{d+2l-4} \\
\label{4.34}
B &=& A + (l-1)\frac{d+2l-6}{d+2l-4} \eta^{l}
\label{4.35}
\end{eqnarray}
We assume that this is the two-point function of the sum of two fields
\begin{equation}
\gamma_{1}\varphi_{1}^{l-1} + \gamma_{2}\varphi_{2}^{l-1}
\label{4.36}
\end{equation}
whose quantum numbers are the same, namely $[\delta(j^{l}) +1, l-1]$, but $\varphi_{1}^{l-1}$ is a new
quasiprimary field whereas $\varphi_{2}^{l-1}$ is a first generation secondary field $a\nabla\tau^{l-2}$. The ansatz that $\tau^{l-2}$ is $j^{l-2,t=1}$ fits well if the anomalous dimensions are equal
\begin{equation}
\eta^{l} = \eta(j^{l-2,t=1})
\label{4.37}
\end{equation}
If this is not the case $\tau^{l-2}$ must be a second new quasiprimary field whose
anomalous dimension is $\eta^{l}$ and is therefore almost degenerate with $j^{l-2,t=1}$. This can only be
decided by a conformal partial wave analysis of an appropriate n-point function.  

The fields $\varphi_{1,2}^{l-1}$ are orthogonal in the Hilbert space due to their different conformal transformation behaviour for equal quantum numbers and we obtain from the two-point function
\begin{eqnarray}
|\gamma_{1}|^{2} &=& 2 l \mathcal{N}_{l}\{ [d+l-1 -2\frac{l-1}{d+2l-4}]\eta^{l} \nonumber \\
                 &+&   [2-\frac{(l-1)(d+2l-6)}{(d+2l-4)(d+2l-4+\eta^{l})}](\eta^{l})^{2}\} \\
\label{4.38}
|\gamma_{2}|^{2} &=& 2l\mathcal{N}_{l}\frac{(l-1)(d+2l-6)}{(d+2l-4)(d+2l-4+\eta^{l})}(\eta^{l})^{2}
\label{4.39}
\end{eqnarray}
We recognize that the secondary field appears at order $(\eta^{l})^{2}$. Redefining the Goldstone boson
by subtraction of the term $2\gamma_{2}a\nabla \tau^{l-2}$ or modifying the de Donder gauge condition (3.8) by such second order renormalization term, leaves only the quasiprimary part for the Goldstone boson. The field $\theta^{l-2}$ appears then as an ancestor field (whose derivative is quasiprimary).

Finally we can translate back the Goldstone field two-point function to $AdS$ space and set $d=3$
\begin{eqnarray}
<G^{l-1}(w_{1};a)G^{l-1}(w_{2};b)> &=& 8\eta^{l}\frac{l^{2}(2l+1)}{2l-1}\mathcal{N}_{l}[(I_{1}-\frac{1}{\zeta}I_{2})^{l-1} \zeta^{-\delta(j^{l})
-1}F_{0}(\zeta) \nonumber\\ &+&
\sum_{k=1}^{l-1}I_{1}^{l-k-1}I_{2}^{k}\zeta^{-\delta(j^{l})-k-2}f_{k}(\zeta)] + O((\eta^{l})^{2})
\label{4.40}
\end{eqnarray}
where $F_{0} = 1$ is fixed. For finite coupling the functions $F_{0}$ and $f_{k}$ have not yet been
calculated.

\setcounter{equation}{0}
\section{Some representation theory background}
Considering representation theory of $SO_{0}(d+1,1)$, i.e. the Euclidean conformal group, the 
intertwining operators acting on C-spaces ($C^{\infty}$ spaces with fixed power behaviour at the infinite
point of $\mathbf{S}^{d}$) of symmetric traceless tensors ("type $1$ representations") 
are all known \cite{Do}. In Section 6B and Propositions 6.1 and 6.2 of that reference we find powers of
gradient and divergence operators 
\begin{equation}
d^{\nu} =(a\nabla)^{\nu} \quad \textnormal{and} \quad (d')^{\nu} = (\frac{\partial}{\partial b}\nabla)^{\nu}
\label{5.1}
\end{equation}
that intertwine exceptional representations of four types
\begin{eqnarray}
\chi_{l\nu}^{\pm} &=& [\pm (d+l+\nu -1), l] \\
\label{5.2}
\chi_{l\nu}^{,\pm} &=& [\pm(d+l-1), l+\nu]
\label{5.3}
\end{eqnarray}
From the quartet diagram of exact sequences of intertwining operators we find in \cite{Do} 
that
\begin{equation}
(d')^{\nu}: C_{l\nu}^{,+} \longrightarrow C_{l\nu}^{+}
\label{5.4}
\end{equation}
Replacing here $l \rightarrow l-1$ and setting $\nu = 1$ we see that the representation
$\chi_{l-1,1}^{,+} = [d+l-2, l]$ is the "conserved current representation" and the divergence maps it
in the representation $ \chi_{l-1,1}^{+} = [d+l-1,l-1]$ which is the "free Goldstone representation".
But the current being conserved must lie in the kernel of this map which from the exact sequence
is the image of the intertwiner $G_{l-1,1}^{,+}$ applied to the space $C_{l-1,1}^{,-}$ or the image of 
the imbedding map $i'_{+}$ applied to the space $D_{l-1,1}^{+}$. In section 7E of \cite{Do} we find that this subspace carries a discrete series representation. 

Such result for the Euclidean conformal group is not final for the physicist: 
We must continue analytically to the Minkowski situation (see \cite{Lu}) and consider the universal covering group of $SO_{0}(d,2)$ to permit irrational anomalous dimensions.  
Then Wightman's axioms leave only a few candidates of the representations for covariance of
fields. In the case of type 1 representations $[\Delta, l\not= 0]$ these are all
\begin{equation}
\Delta \geq d+l-2
\label{5.5}   
\end{equation}
and the limiting case $\Delta = d+l-2$ are the conserved currents. 
On the other hand we have analyzed the divergence of a two-point function. This implies by Wightman's axioms that we consider very special representations of the universal covering group, namely representations carried by spaces of holomorphic functions on the forward tube in complex Minkowski space. To obtain them from C-spaces we can apply two-point functions as intertwiners on the C-spaces and 
impose analyticity on them this way \cite{R2}, \cite{Ma}.  

We have considered the deformed case in this work. In this case the divergence of the current is not zero
and an anomalous dimension appears. Then the conservation is replaced by
\begin{equation}
d'j^{l}(x; a) = \gamma_{1}\varphi_{1}^{l-1} + \gamma_{2}\varphi_{2}^{l-1}
\label{5.6}
\end{equation}
which we can consider as definitions of new quasiprimary fields.


\begin{thebibliography}{99}
\bibitem{Kl}
I.~R.~Klebanov, A.~M.~Polyakov: "AdS dual of the critical $O(N)$ vector model", Phys.
Letters B550 (2002) 213-21 [arXiv hep-th/0210114]
\bibitem{Le}
T.~Leonhardt, A.~Meziane, W.~R\"uhl: "On the proposed AdS dual of the critical $O(N)$ sigma model for any dimension $2<d<4$", Physics Letters B555 (2003) 271-278 [arXiv:hep-th/0211092]
\bibitem{LR}
K.~Lang, W.~R\"uhl:"The critical O(N) $\sigma$-model at dimensions $2<d<4$: Hardy-Ramanujan distribution of quasi-primary fields and a collective fusion approach", Z. Phys. C 63 (1994) 531-539
\bibitem{R1}
W.~R\"uhl: "The masses of gauge fields in higher spin field theory on AdS(4)", Phys. Letters B605 (2005)
413-418 [arXiv:hep-th/0409252]
\bibitem{MR1}
R.~Manvelyan, W.~R\"uhl: "The off-shell behavior of propagators and the Goldstone field in higher spin gauge theory on $AdS_{d+1}$ space", Nucl. Phys. B717 (2005) 3-18 [arXiv:hep-th/0502123]  
\bibitem{MR2}
R.~Manvelyan, W.~R\"uhl: "The masses of gauge fields in higher spin field theory on the bulk of $AdS_{4}$",
Physics Letters B613 (2005) 197-207 [arXiv:hep-th/0412252]
\bibitem{Do}
V.~K.~Dobrev, G.~Mack,V.~B.~Petkova, S.~G.~Petrova, I.~T.~Todorov: "Harmonic analysis on the n-dimensional Lorentz group and its application to conformal quantum field theory", Lecture Notes in Physics vol. 63,
Springer 1977
\bibitem{Lu}
M.~L\"uscher, G.~Mack: "Global conformal invariance in quantum field theory", Commun. math. Phys. 41 (1975) 203-234
\bibitem{R2}
W.~R\"uhl: "Field representations of the conformal group with continuous mass spectrum", Commun. math. Phys. 30 (1973) 287;  "On conformal invariance of interacting fields", Commun. math. Phys. 34 (1973) 149
\bibitem{Ma}
G.~Mack: "All unitary ray representations of the conformal group SU(2,2) with positive energy", Commun. 
math. Phys. 55 (1977) 1-28 (this is only for $d = 4$ whereas we are mainly interested in $d = 3$)
\bibitem{Vas}
A.N.~Vasiliev, Y.M.~Pismak and Y.R.~Khonkonen: Teor. Mat. Fiz. 46 (1981) 157; "Simple method of calculating the critical indices in the 1/N expansion", Theor. Math. Phys. 46 (1981) 104 (Engl. transl.) 
\bibitem{RL}
K.~Lang and W.~Rühl: "The critical O(N) $\sigma$-model at dimensions $2<d<4$: a list of quasi-primary fields", Nucl. Phys. B402 (1993) 573; "The critical O(N) $\sigma$-model at dimension $2<d<4$: Fusion coefficients and anomalous dimensions", Nucl. Phys. B400 (1993) 597
\bibitem{RG}
I.S.~Gradshteyn and I.M.~Ryzhik, "Table of Integrals, Series, and Products", eq. 8.932.1, New York (1965)
\bibitem{F}
C.~Fronsdal: "Singletons and massless, integral spin fields on de Sitter space (elementary particles in curved space vii)", Phys. Rev D20 (1979) 848; "Massless fields with integer spin", Phys. Rev D18 (1978) 3624  


\end{thebibliography}
\end{document}